# Visual anemometry of natural vegetation from their leaf motion

Roni H. Goldshmid,[1,2,5*] John O. Dabiri[1,3*] and John E. Sader[1,4*]

[1] Graduate Aerospace Laboratories, California Institute of Technology, Pasadena CA 91125, USA

[2] Department of Aerospace Engineering, San Diego State University, San Diego CA 92182, USA

[3] Department of Mechanical and Civil Engineering, California Institute of Technology, Pasadena CA 91125, USA

[4] Department of Applied Physics and Materials Science, California Institute of Technology, Pasadena CA 91125, USA

[5] Present address: Department of Aerospace Engineering, San Diego State University, San Diego, CA 92182, USA.

* rgoldshmid@sdsu.edu, jodabiri@caltech.edu, jsader@caltech.edu

High-resolution, near-ground wind-speed data are critical for improving the accuracy of weather predictions and climate models,[1–3] supporting wildfire control efforts,[4–7] and ensuring the safe passage of airplanes during takeoff and landing maneouvers.[8,9] Quantitative wind speed anemometry generally employs on-site instrumentation for accurate single-position data or sophisticated remote techniques such as Doppler radar for quantitative field measurements. It is widely recognized that the wind-induced motion of vegetation depends in a complex manner on their structure and mechanical properties, obviating their use in quantitative anemometry.[10–14] We analyze measurements on a host of different vegetation showing that leaf motion can be decoupled from the leaf's branch and support structure, at low-to-moderate wind speed, $U_{\text{wind}}$. This wind speed range is characterized by a leaf Reynolds number, enabling the development of a remote, quantitative anemometry method based on the formula, $U_{\text{wind}} \approx 740\sqrt{\mu U_{\text{leaf}}/\rho D}$, that relies only on the leaf size $D$, its measured fluctuating (RMS) speed $U_{\text{leaf}}$, the air viscosity $\mu$, and its mass density $\rho$. This formula is corroborated by a first-principles model and validated using a host of laboratory and field tests on diverse vegetation types, ranging from oak, olive, and magnolia trees through to camphor and bullgrass. The findings of this study open the door to a new paradigm in anemometry, using natural vegetation to enable remote and rapid quantitative field measurements at global locations with minimal cost.

The attainment of ground-level wind-speed maps involves a fundamental tradeoff between the sparse coverage provided by on-site instrumentation, and the cost and limited mobility of deploying remote sensing techniques such as Doppler radar and LIDAR. In contrast, the technique proposed by Admiral Sir Francis Beaufort—commonly referred to as the Beaufort Scale—requires no such instrumentation and instead relies on visual cues from natural observations to estimate the wind strength. The wind strength is reported on a scale from 0 to 12, ranging from calm to hurricane force. Combining the simplicity of the Beaufort Scale with the quantitative capabilities of modern anemometry would facilitate the broad deployment of anemometry on a global scale.

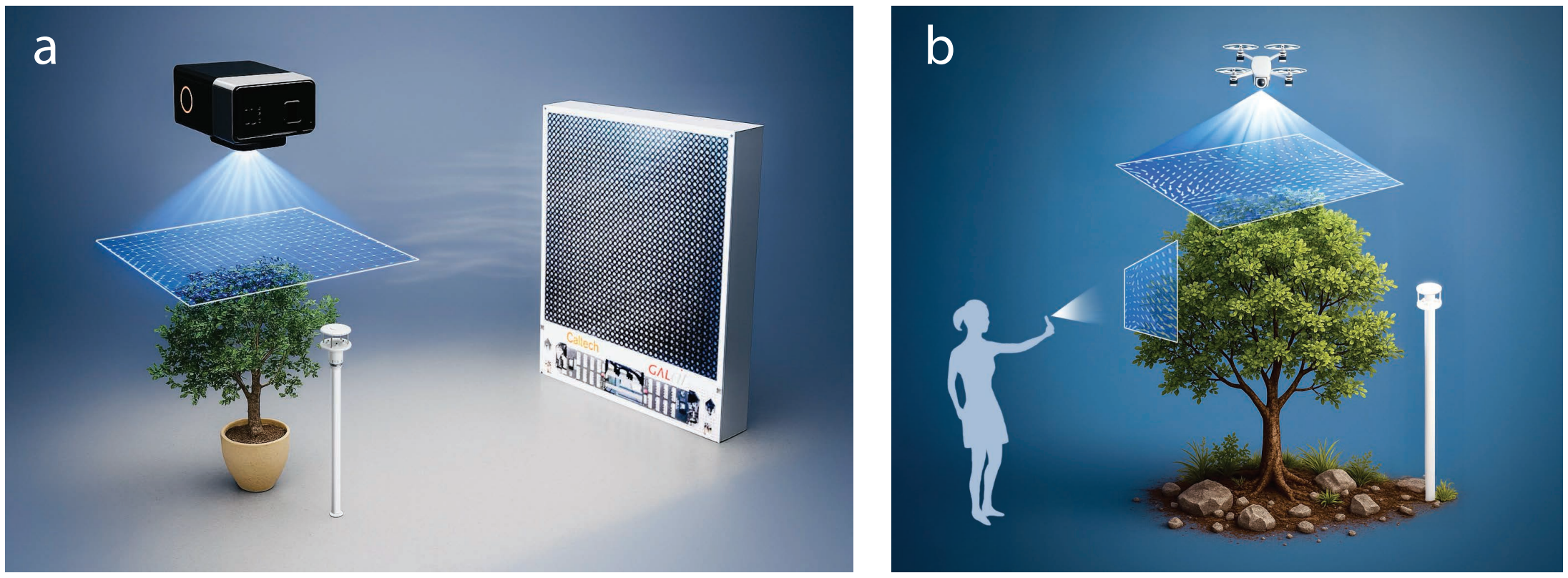

**Fig. 1 | Visual anemometry. a**. Protocol for the wind tunnel measurements on the vegetation species reported in Ref. 15. A precision fan array (right) generates the impinging wind on the subject plant (left), which is measured using a conventional anemometer (adjacent to the plant). Plant motion is collected using a camera (overhead). Details of the measurement are in Methods. **b.** Field measurements using visual anemometry illustrating its potential use cases with a ground-based mobile phone (person) or a remote drone (overhead). An instrumented anemometer (right) is next to the plant, illustrating conventional wind speed measurements. Image credit: Peakcells LLC.

Visual anemometry aims to provide this capability, drawing upon the natural motion of plants in the wind to quantitatively estimate the wind speed.[14] Existing visual anemometry methods combine video footage of vegetation motion with machine learning[10,13] and physics-based models[11–13] to yield wind speed estimates. However, these models exhibit a strong dependence on the structural properties of the vegetation, requiring independent calibration of each plant specimen using mathematical modeling or a separate instrumented anemometer. This calibration must be updated continuously in time on a plant-by-plant basis, due to plant change and growth, severely limiting the utility of this approach and its remote deployment in the field.

We examine the foundations of visual anemometry and discover a previously unforeseen phenomenon: *the motions of leaves can be decoupled from their supporting branch structure at low-to-moderate wind speeds*. That is, leaves can behave as natural "particles" that are held in position by the plant's support structure (branches). Leaf motion can then be analyzed to infer the impinging wind speed—a form of natural particle tracking velocimetry. This finding enables the development of a physics-based visual anemometry technique that does not rely on the support structure, e.g., branch stiffness and size. It enables, for the first time, the quantitative measurement of wind speed from the remote observation of plant motion.

An online database,[15] that reports precision wind tunnel measurements by two of the authors on a diverse range of vegetation species, is used for the initial analysis. The database consists of one grass and seven tree species: bullgrass, camphor, mesquite, oak, olive, paperbark, pepper, and pine (Methods). Measurements are reported in the wind speed range of 1 to 15 m/s. The RMS leaf speed is measured by analyzing successive images taken from videography of each plant. The measurement protocol detailed in Ref. 15 is illustrated in Fig. 1(a) and summarized in Methods. Figure 2(a) reports key data from this online database, showing that (1) the measured root-mean-square (RMS) leaf speed generally increases with the impinging wind speed, and (2) there is no apparent trend connecting the measured leaf speed to the different plant species. The leaves of all plants exhibit time-dependent fluctuations about an average position, i.e., the RMS leaf speed is nonzero, due to the impinging wind speed and the restraining effect of their branch structure.

Flow instabilities generated by a solid body in a uniform stream are typically initiated at a Reynolds number of $\mathrm{Re} \approx 50$ to 100, with transition in the shed free shear layers as Re approaches and exceeds $\approx 1{,}000$.[16] Calculating a Reynolds number, $\mathrm{Re}_{\mathrm{leaf}}$, based on the leaf size (Methods) for the dataset reported in Fig. 2(a) reveals $1{,}000 \lesssim \mathrm{Re}_{\mathrm{leaf}} \lesssim 30{,}000$. This immediately suggests that local flow instabilities generated by individual leaves may be responsible for their observed fluctuating motions in a tree.

To explore this possibility, the dataset in Fig. 2(a) is analyzed using dimensional analysis.[17] Dependence of the fluctuating leaf speed, $U_{\mathrm{leaf}}$, on the leaf size, $D$, the impinging wind speed, $U_{\mathrm{wind}}$, together with the air viscosity, $\mu$, and air mass density, $\rho$, is sought. This represents a minimal set of dependent variables for this problem. The analysis proceeds by requiring the constants, $\alpha_0$, $\alpha_1$, $\alpha_2$, and $\alpha_3$, such that the monomial of all variables, i.e., $U_{\text{leaf}}\, U_{\text{wind}}^{\alpha_0} D^{\alpha_1} \mu^{\alpha_2} \rho^{\alpha_3}$, is dimensionless. The general solution is $\alpha_1 = -\alpha_2 = \alpha_3 = 1 + \alpha_0$, from which the specified monomial becomes: $\rho U_{\text{leaf}} D/\mu\, (\rho U_{\text{wind}} D/\mu)^{\alpha_0}$. This establishes that two dimensionless groups exist:

$$\text{Re}_{\text{leaf}} = \frac{\rho U_{\text{leaf}} D}{\mu}, \qquad \text{Re}_{\text{wind}} = \frac{\rho U_{\text{wind}} D}{\mu}, \tag{1}$$

corresponding to a Reynolds number for the fluctuating leaf speed (as defined above) and another for the impinging wind speed; both referenced to the leaf size. Application of Eq. (1) to the horizontal and vertical axes in Fig. 2(a)—which involves a trivial multiplication by $\rho D/\mu$—produces a remarkable and previously unforeseen collapse of the measured dataset in Ref. 15. This collapse is observed regardless of the plant species and their structural properties. This universal data collapse, over the entire measurement range, strongly supports the aforementioned "independent leaf hypothesis" that the observed leaf motion is due to flow instabilities generated by the leaf itself.

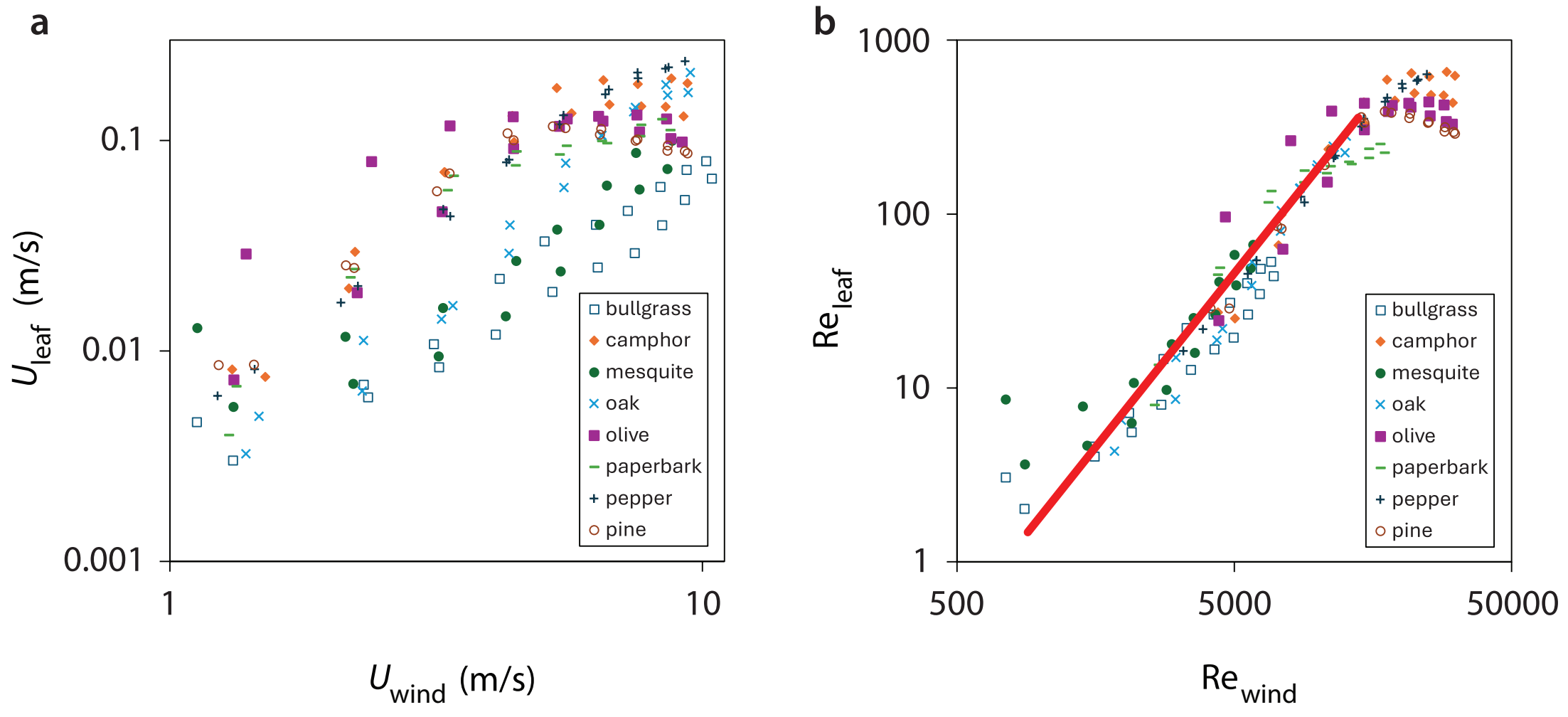

**Fig. 2 | Leaf speed versus wind speed and the non-dimensionalization for visual anemometry.** **a.** Measured data[15] plotted on a double-logarithmic scale that exhibits one decade of variation in leaf speed at each wind speed. **b.** The non-dimensionalizaton in Eq. (1) collapses the measured data over the entire measurement range. Variation along the vertical axis in **b** for each specified value on the horizontal axis is greatly reduced from its dimensional counterpart in **a**. The red line, $\text{Re}_{\text{leaf}} =$

$(\mathrm{Re}_{\mathrm{wind}}/740)^2$, is a best fit of the measured dataset to a squared power-law dependence of $\mathrm{Re}_{\mathrm{leaf}}$ on $\mathrm{Re}_{\mathrm{wind}}$.

The dimensionless dataset in Fig. 2(b) exhibits a clear functional dependence of $\mathrm{Re}_{\mathrm{leaf}}$ on $\mathrm{Re}_{\mathrm{wind}}$, for $\mathrm{Re}_{\mathrm{wind}} \lesssim 10{,}000$, i.e., $\mathrm{Re}_{\mathrm{leaf}} = f(\mathrm{Re}_{\mathrm{wind}})$. This functional form is anticipated by the Buckingham Pi theorem[17] provided the principal dimensional variables are included in the dimensional analysis. Least squares fit of the dataset for $1{,}000 \leq \mathrm{Re}_{\mathrm{wind}} \lesssim 10{,}000$ to a quadratic power law, where the functional dependence of $\mathrm{Re}_{\mathrm{leaf}}$ on $\mathrm{Re}_{\mathrm{wind}}$ is non-constant, reveals the empirical form, $\mathrm{Re}_{\mathrm{leaf}} \approx (\mathrm{Re}_{\mathrm{wind}}/a)^2$, where $a = 740 \pm 30$ (95% C.I.). This provides an accurate predictor of the dataset over two decades in $\mathrm{Re}_{\mathrm{leaf}}$. Rearranging this expression and utilizing Eq. (1) then gives the dimensional formula,

$$U_{\mathrm{wind}} \approx 740 \sqrt{\frac{\mu U_{\mathrm{leaf}}}{\rho D}}, \qquad 1{,}000 \lesssim \mathrm{Re}_{\mathrm{wind}} \lesssim 10{,}000. \tag{2}$$

Equation (2) enables the wind speed, $U_{\mathrm{wind}}$, impinging on a plant to be measured from remote observations of the fluctuating motion (RMS speed) of its leaves, $U_{\mathrm{leaf}}$.

Fully developed turbulent flow around a solid body is commonly generated at Reynolds numbers exceeding $\approx 10^6$.[16] This value well exceeds the measurement range reported in Fig. 2. The visual anemometry method embodied in Eq. (2) thus applies at low-to-moderate wind speed, as characterized by the impinging wind Reynolds number, $\mathrm{Re}_{\mathrm{wind}}$. The dataset in Fig. 2(b) also shows that $\mathrm{Re}_{\mathrm{leaf}}$ is approximately constant for $\mathrm{Re}_{\mathrm{wind}} \gtrsim 10{,}000$. This observation aligns with literature reports of the fluctuating drag coefficient experienced by circular cylinders.[16] Despite the difference in geometry between a cylinder and a leaf, this similitude provides further support for the articulated "independent leaf hypothesis" which underlies Eq. (2). However, it indicates that visual anemometry by the present method may not be possible for $\mathrm{Re}_{\mathrm{wind}} \gtrsim 10{,}000$. Plant motion in this high-speed regime is also characterized by a collective downstream sweeping of the branches and leaves (Supplementary Information). Leaf-leaf and branch-branch interactions, along with branch stiffness, are expected to play an increasingly dominant role in this high-speed regime.

The numerical factor of 740 in Eq. (2) appears unusual at first sight but has a simple physical origin. This is explained through the development of a first-principles theoretical model. We draw upon literature data for the fluctuating drag coefficient experienced by a cylinder of diameter $D$ with an impinging flow Reynolds number of $\mathrm{Re_{wind}} \approx 11{,}000$ (referenced to the cylinder diameter).[18] Solving the equation of motion for a free bluff body which is (1) restrained from moving with the average downstream flow, and (2) subjected to this fluctuating hydrodynamic load, gives $\mathrm{Re_{wind}}/\sqrt{\mathrm{Re_{body}}} \approx 700$ where $\mathrm{Re_{body}} = \rho U_{\mathrm{body}} D/\mu$ and $U_{\mathrm{body}}$ is the RMS fluctuating body speed (Methods). This independent calculation, on a related but different system, agrees remarkably well with the empirical expression for the eight plants in Ref. 15, from which Eq. (2) is derived. It suggests that the physical origin of the factor of 740 in Eq. (2) is rooted in the fluctuating drag coefficient experienced by the body. This hydrodynamic load is generated intrinsically by local flow instabilities. This first-principles calculation thus provides further evidence in support of the hypothesis that local fluctuating hydrodynamic forces drive the leaf motion in natural vegetation.

We assess the robustness of the discovered natural anemometry formula, Eq. (2), using field measurements of different trees on the Caltech campus. Figure 3 gives data obtained using remote observations of two trees: a California sycamore (*platanus racemosa*) and an Engelmann oak (*quercus engelmannii*). A sonic anemometer is placed next to each tree and used to independently measure the local wind speed as a function of time. Videography from the side of the tree is analyzed using the protocol in Methods to yield time series of the RMS leaf speed, $U_{\mathrm{leaf}}$; as in Fig. 1(b) but using a laboratory camera. The instantaneous wind speed, $U_{\mathrm{wind}}$, is recovered from this dataset using Eq. (2) (VA) and compared with the independent sonic anemometry measurements (sonic) in Fig. 3. Quantitative agreement between the two independent measurements is observed. with the mean difference of all measured wind speeds of the sycamore and oak trees being $0.07 \pm 0.27$ (SD) m/s and $0.01 \pm 0.33$ (SD) m/s, respectively; see Fig. 3(a,b). Leaf size uncertainty is halved in Eq. (2) due to its square root dependency, leading to an improved wind speed estimate. Time series of the instantaneous measured wind speeds for datasets displaying the highest cross correlation coefficient, $\mathrm{cov}\big(U_{\mathrm{wind}}^{\mathrm{VA}}, U_{\mathrm{wind}}^{\mathrm{sonic}}\big)/\big(\mathrm{SD}\big(U_{\mathrm{wind}}^{\mathrm{VA}}\big)\mathrm{SD}\big(U_{\mathrm{wind}}^{\mathrm{sonic}}\big)\big)$ between the sonic anemometer and Eq. (2), are given in Figs. 3(c,d); datasets with the lowest cross correlation coefficients are in Figs. 3(e,f). Times series for all datasets are reported in the Supplementary Information.

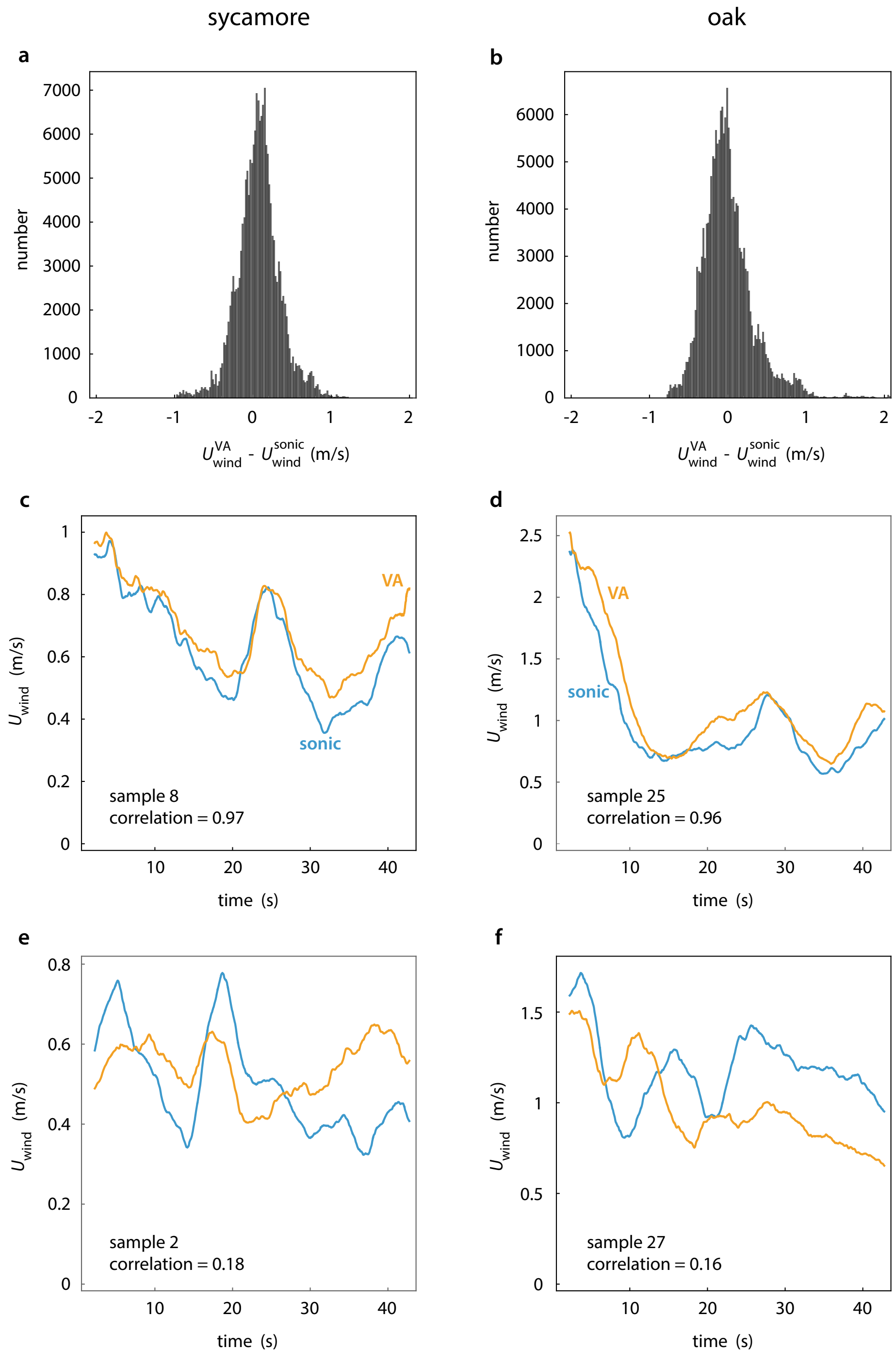

**Fig. 3 | Application of visual anemometry in the field.** Field measurements taken using a California sycamore (platanus racemosa) [left] and an Engelmann oak (quercus engelmannii) [right]. Benchmark wind speeds are measured using a sonic anemometer placed next to the trees. **a, b.** Histograms of the wind speed difference in using Eq. (2) relative to the sonic anemometer. **c, d.** Measured time series that exhibit the highest cross correlation coefficients between Eq. (2) and the

sonic anemometer. **e, f.** Time series exhibiting the lowest correlation. Sonic and VA refer to the sonic anemometer and Eq. (2), respectively.

Importantly, the sonic anemometer measures the wind speed at a single position while the visual anemometry method provides a tree-averaged measurement. Wind speed variability across each tree may therefore be a complicating factor in the reported comparison, in contrast to the wind tunnel measurement in Ref. 15. Even so, the data reported in Fig. 3 provides a practical validation of Eq. (2) in a natural setting and displays the capabilities of Eq. (2) in accurately measuring the wind speed.

A quantitative link between the natural motion of vegetation and the impinging wind speed has been found. This was achieved by identifying the principal physical mechanism underlying leaf motion: local hydrodynamic flow instabilities are driven by each leaf. This finding is distinct from previous studies that have focused on the overall motion of trees that must intrinsically depend on the mechanical properties of their supporting branch structures.[11–13] The discovered formula in Eq. (2)—that primarily depends on the leaf size—enables remote and quantitative natural anemometry on the ground, using portable devices such as mobile phones, through to low-earth-orbit satellite videography; Fig. 1(b). The latter currently provides commercial video feeds at $\approx$ 60 frames per second that could be used for visual anemometry, albeit at half-meter resolution. This falls short of the resolution achievable with aerial imaging systems. The required resolution to measure leaf motion may soon be available using advanced satellite videography, enabling the broad deployment of anemometry on a global scale.

# Methods

## Wind tunnel measurements

Measurements in Ref. 15 used the open circuit wind tunnel at the Caltech Center for Autonomous Systems and Technologies (CAST). Vegetation kinematics were recorded using an overhead camera and leaf fluctuating speeds estimated using cross correlations (next section). The ground truth wind speed was recorded using a sonic anemometer installed adjacent to the vegetation. Seven species of trees and a patch of grass were investigated, which ranged in height from 1 to 4 meters with their widths in the range of 1.5 to 3 meters. Each of the plants was exposed to wind speeds ranging from 1 to 15 m/s. Leaf sizes are reported below.

## Protocol for measuring the leaf speed

The RMS leaf speed of each plant reported in this study is measured using a small control volume (areal region in the video) that is applied in a patchwork manner across the plant to provide full canopy coverage. Cross-correlation coefficients between successive video frames of each control volume are computed using PIVlab[19] in MATLAB. This yields the RMS displacement vector of all leaves within each control volume, and hence the RMS displacement field across the plant from all control volumes; length is calibrated using a reference object. Two windows are used to sample each video frame of each control volume, from which the cross-correlation coefficients are computed: (1) a 64-pixel by 64-pixel window with a step size of 32 pixels, and (2) a 32-pixel by 32-pixel window with a step size of 16 pixels; these yield consistent results. The velocity field is calculated from this RMS displacement field using the known frame rate of the camera (60 Hz). The magnitude of the velocity vector in each control volume is averaged across all control volumes to yield the RMS leaf speed for each plant reported in this study. Time series of the RMS leaf speed then follows by applying this analysis to all successive video frames. Individual leaves are not tracked using this method, but the speed across all leaves in each control volume is measured.

## Leaf sizes

Measured leaf sizes, $D$, in the following table are used to non-dimensionalize the data in Fig. 2(a) to obtain Fig. 2(b). These were obtained by averaging the size of several leaves in a single plant, and the data are reported to measurement precision. Data are leaf lengths, unless stated otherwise.

| Plant | Leaf size (cm) |
|---|---|
| Muhlenbergia emersleyi (Bullgrass) | 1 (width) |
| Cinnamomum camphora (Camphor tree) | 5 |
| Prosopis alba thornless (Mesquite tree) | 1 |
| Quercus agrifolia (Oak tree) | 2 |
| Olea europaea (Olive tree) | 5 |
| Melaleuca quinquenervia (Paperbark tree) | 3 |
| Schinus molle (Pepper tree) | 4 |
| Pinus radiata (Pine tree) | 5 |
| Platanus racemosa (Sycamore tree, Caltech) | 14 |
| Quercus engelmannii (Oak tree, Caltech) | 3 |

## First-principles model

A first-principles calculation is reported that connects the impinging wind speed, $U_{\text{wind}}$, to the RMS leaf speed, $U_{\text{leaf}}$. This models the leaf as a 2D bluff body and draws upon independent experimental data for the fluctuating drag force that a cylinder experiences in an impinging uniform flow. As for the leaf, the bluff body is restrained from moving downstream with the mean flow. The equation of motion for the fluctuating kinematics of the body is

$$\frac{d}{dt}\left[\left(M_{\text{body}} + M_{\text{added}}\right) U(t)\right] = F_{\text{fluc}}(t), \tag{3}$$

where $t$ is time, $M_{\text{body}}$ is the body mass per unit length, and $M_{\text{added}} = (\pi/4)\rho D^2$ is the added mass per unit length due to the surrounding fluid, respectively, $\rho$ is the fluid density, $D$ is the body diameter, $U(t)$ is the instantaneous velocity of the body in the downstream direction, and $F_{\text{fluc}}(t)$ is the fluctuating drag force per unit length in the downstream direction generated by the body. Independent measurements of the fluctuating drag force generated by a cylinder in a uniform flow of speed, $U_{\text{wind}}$, have been reported previously.[18] These measurements show that for a Reynolds number of

$$\text{Re}_{\text{wind}} \equiv \frac{\rho D U_{\text{wind}}}{\mu} = 11{,}000, \tag{4}$$

where $\mu$ is the fluid viscosity, an RMS fluctuating drag coefficient per unit length of

$$\Delta C_D \equiv \frac{F_{\text{RMS}}}{\frac{1}{2}\rho D U_{\text{wind}}^2} \approx 0.075, \tag{5}$$

results, with a Strouhal number of

$$\text{St} \equiv \frac{\omega D}{U_{\text{wind}}} \approx 2.5, \tag{6}$$

where $\omega$ is the dominant angular frequency of the fluctuating drag force. Equation (3) is now expressed as

$$\omega M_{\text{added}} U_{\text{body}} = F_{\text{RMS}}, \tag{7}$$

where $U_{\mathrm{body}}$ is the RMS body speed and we have assumed that $M_{\mathrm{body}} \ll M_{\mathrm{added}}$, which is not an unreasonable approximation for a leaf whose plan-view dimensions far exceed its thickness. Substituting Eqs. (5) and (6) into Eq. (7) leads to

$$\left(\frac{2.5\, U_{\mathrm{wind}}}{D}\right)\left(\frac{\pi}{4}\rho D^2\right) U_{\mathrm{body}} \approx \frac{1}{2}\rho U_{\mathrm{wind}}^2 D\, \Delta C_D(\mathrm{Re}_{\mathrm{wind}}), \tag{8}$$

giving the relation

$$U_{\mathrm{wind}} \approx 50\, U_{\mathrm{body}}, \tag{9}$$

i.e., $\mathrm{Re}_{\mathrm{wind}} \approx 50\, \mathrm{Re}_{\mathrm{body}}$ where $\mathrm{Re}_{\mathrm{body}} \equiv \rho D U_{\mathrm{body}}/\mu$. Equation (9) immediately gives the required result,

$$\frac{\mathrm{Re}_{\mathrm{wind}}}{\sqrt{\mathrm{Re}_{\mathrm{body}}}} \approx 700. \tag{10}$$

The first-principles calculation, Eq. (10), agrees with independent leaf measurements leading to Eq. (2).

## Acknowledgements

We thank Peakcells LLC for creating Fig. 1, and acknowledge funding from a Caltech Presidential Postdoctoral Fellowship, the Gary Clinard Innovation Fund, Heliogen Inc, the Center for Autonomous Systems and Technologies at Caltech, and the National Science Foundation (Grant CBET-2019712).

## Author contributions

R.H.G. and J.E.S. formulated the quantitative approach; R.H.G. analyzed experiments and J.E.S. performed theory; R.H.G. and J.O.D. conducted field measurements; all authors discussed theory, measurements and analysis; R.H.G. and J.E.S. wrote the initial draft of the manuscript, and all authors contributed to manuscript revisions.

## Competing interests

The authors declare no competing interests.

S U P P L E M E N T A R Y   I N F O R M A T I O N

# Visual anemometry of natural vegetation from their leaf motion

Roni H. Goldshmid,[1,2,5*] John O. Dabiri[1,3*] and John E. Sader[1,4*]

[1] Graduate Aerospace Laboratories, California Institute of Technology, Pasadena CA 91125, USA

[2] Department of Aerospace Engineering, San Diego State University, San Diego CA 92182, USA

[3] Department of Mechanical and Civil Engineering, California Institute of Technology, Pasadena CA 91125, USA

[4] Department of Applied Physics and Materials Science, California Institute of Technology, Pasadena CA 91125, USA

[5] Present address: Department of Aerospace Engineering, San Diego State University, San Diego, CA 92182, USA.

* rgoldshmid@sdsu.edu, jodabiri@caltech.edu, jsader@caltech.edu

## CONTENTS

## Section A – Images showing downstream sweeping of plants from Ref. 15

At low wind speeds, the leaves of each tree are observed to move independently (Section B). This behavior changes as the wind speed increases with a collective down sweeping of the branches and leaves. Figure S1 provides overhead images of olive and camphor trees at a high wind speed. The impinging wind is directed left-to-right in the image.

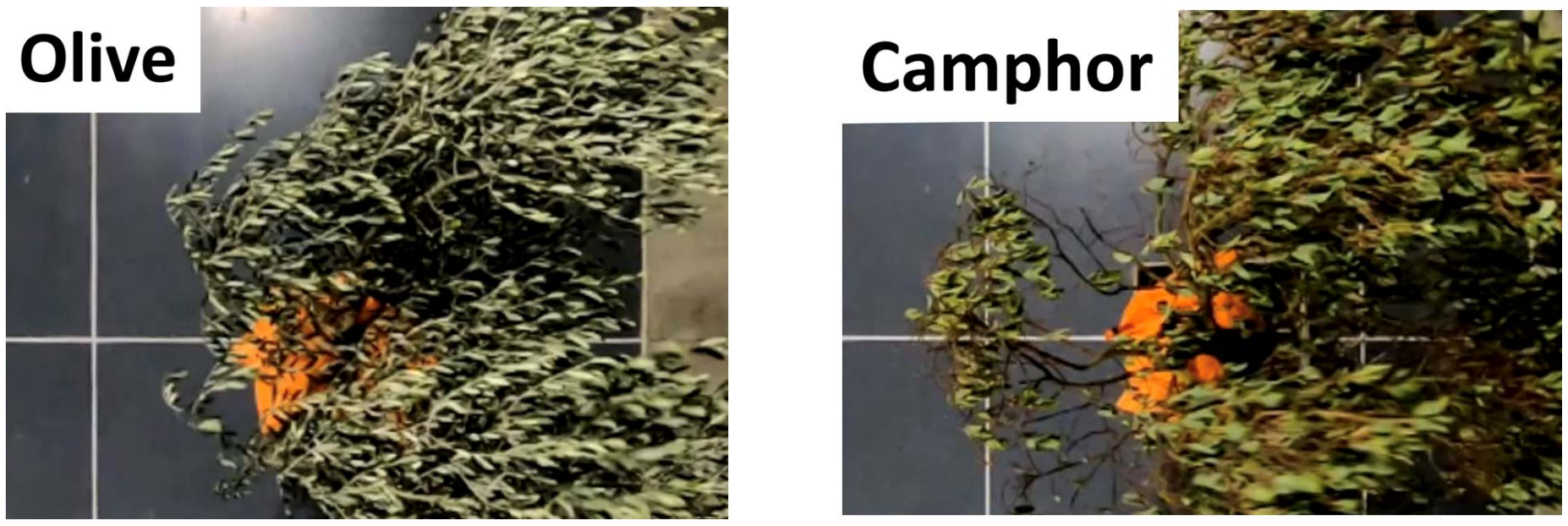

**Figure S1 |** Overhead images of *quercus agrifolia* (oak tree) and *cinnamomum camphora* (camphor tree) at a high wind speed of 10 m/s, showing a collective downstream sweeping of the branches and leaves.

## Section B – Videos demonstrating independent leaf motion

The supplementary videos show two trees whose leaves are moving due to the impinging wind. The first is a camphor tree and the second is an abele tree. Branch motion is also observed but this motion is slower than that of the leaves. This has the effect of reducing the branch contribution to the maximal velocities measured using the PIV software, which predominantly picks up the leaf motion.

## Section C – Time series comparisons of all measurements on a sycamore and an oak tree.

Separate time measurements of 45 sec duration taken on a sycamore and oak tree on the Caltech campus. Time series in Secs. C1, C2 are smoothed using a moving average involved 10% of the total number of points. Unsmoothed data is in Secs. C3, C4 and exhibit identical features and speed magnitudes, with significant noise. Cross correlation coefficients are included at the top of each plot.

C1. Sycamore tree time series (smoothed – moving average); sonic (blue), VA (orange)

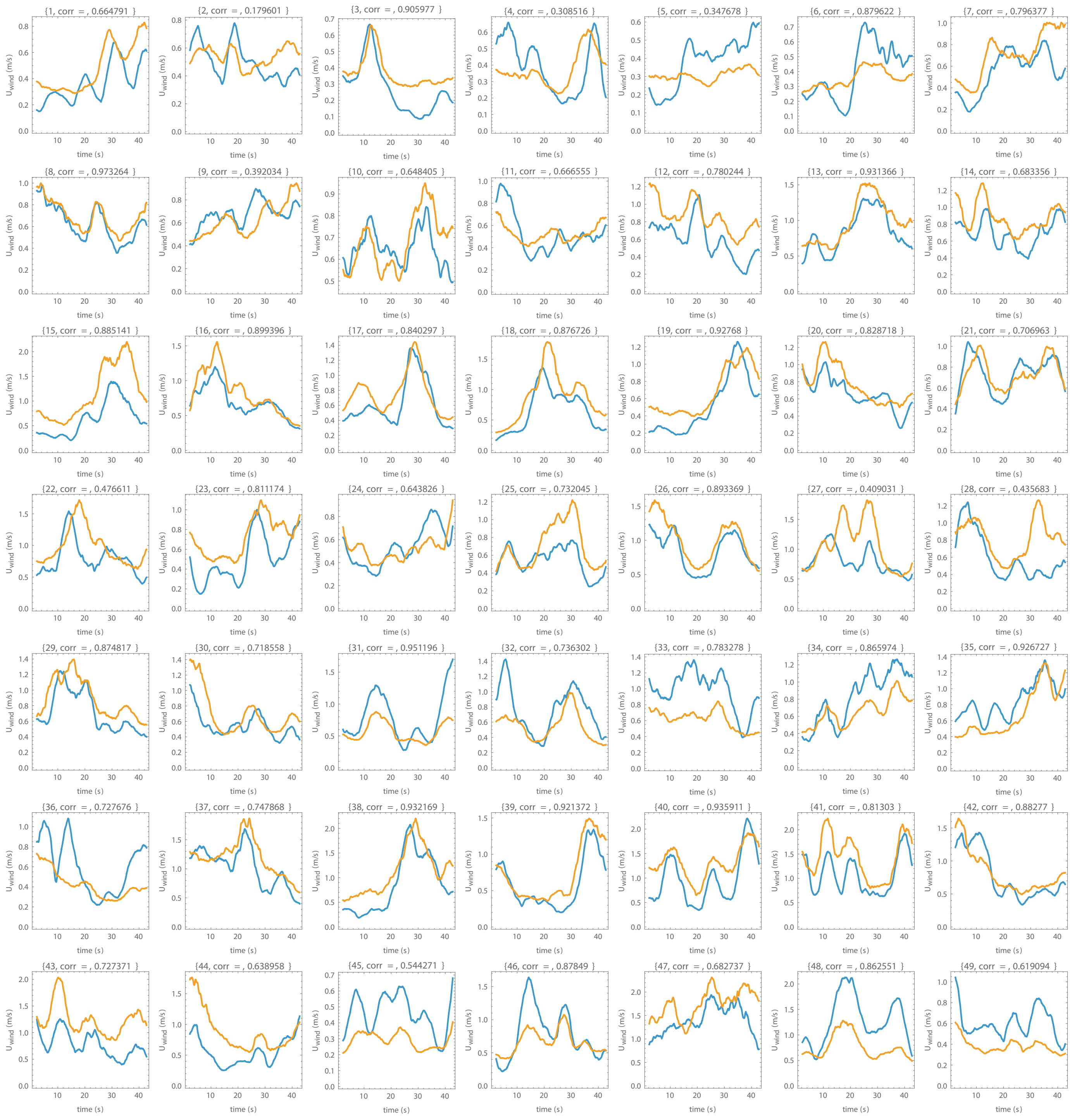

# C2. Oak tree time series (smoothed – moving average); sonic (blue), VA (orange)

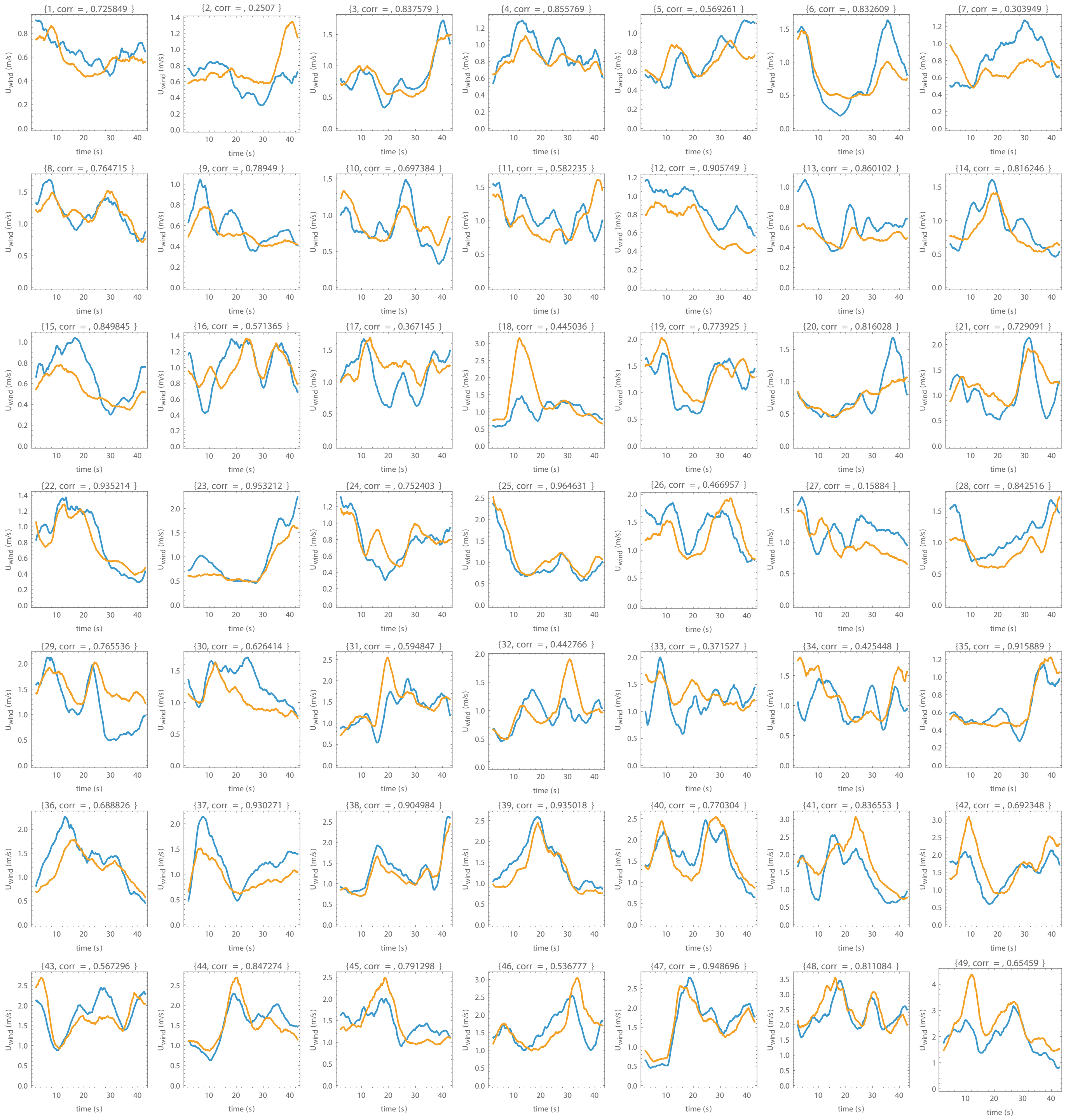

## C3. Sycamore tree time series (raw data); sonic (blue), VA (orange)

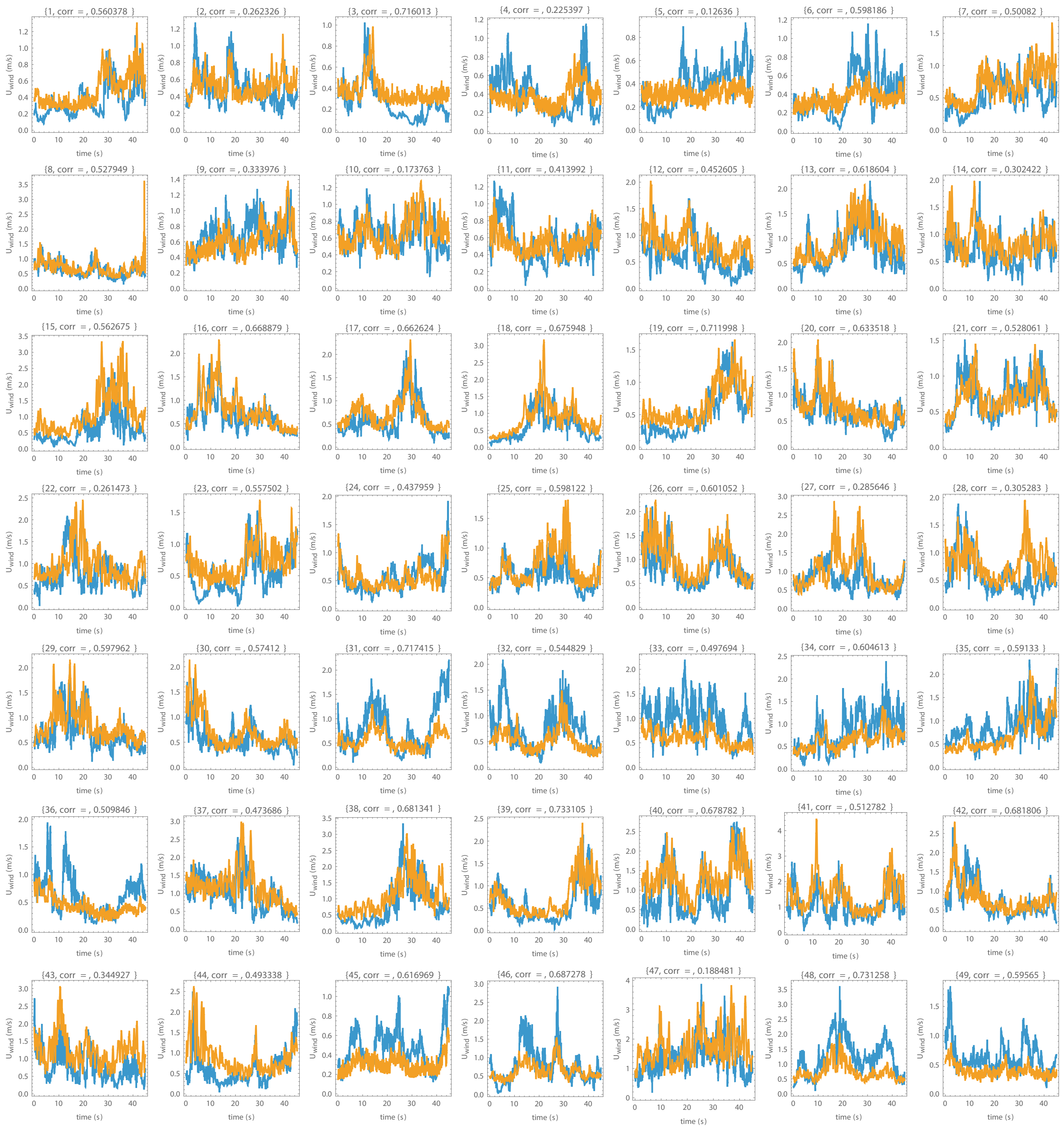

# C4. Oak tree time series (raw data); sonic (blue), VA (orange)

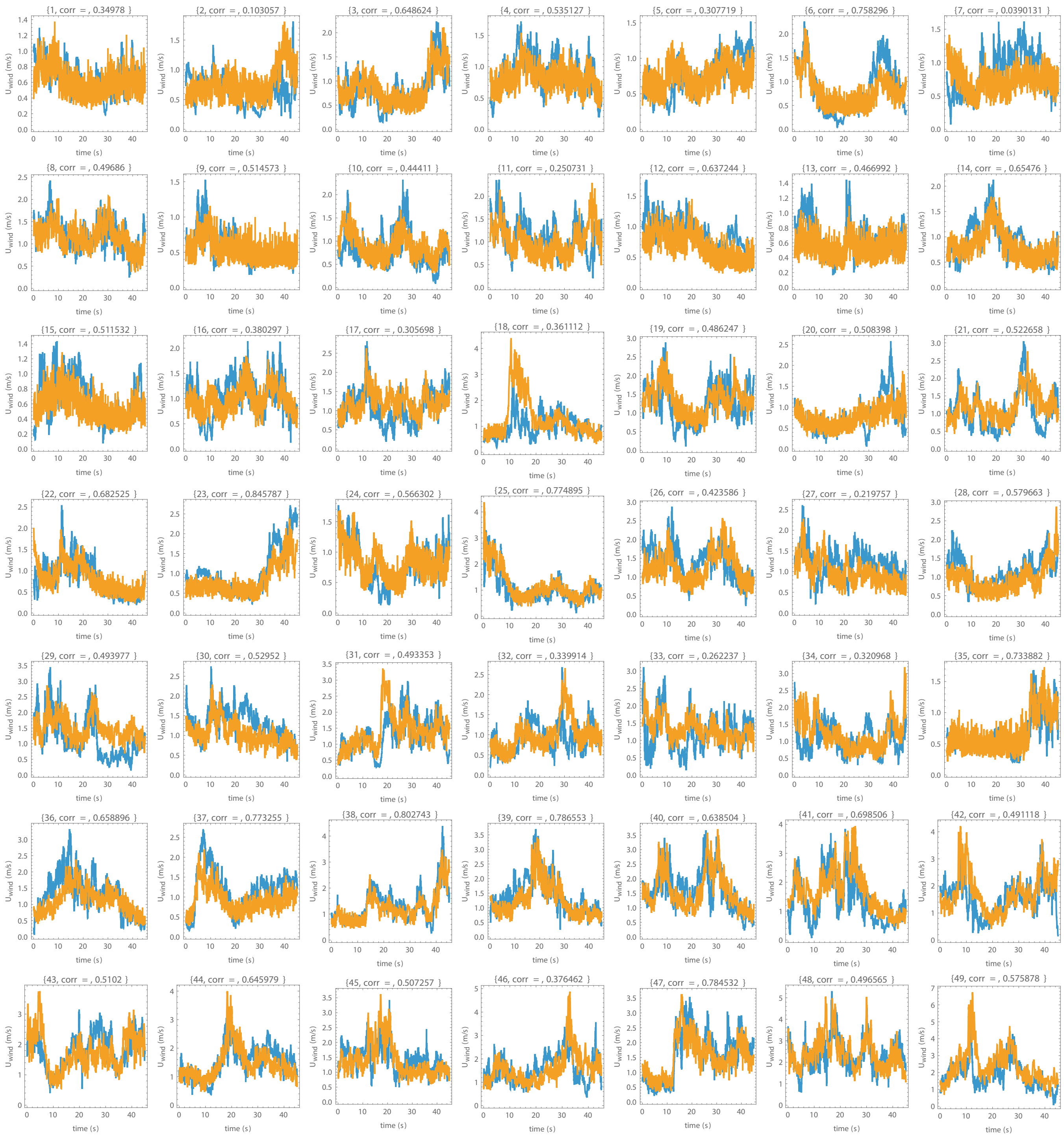

C5. Histograms of wind speed differences between sonic anemometer and Eq. (2) (raw data)

Histogram of the difference in the wind speed measurement between the sonic anemometer and Eq. (2), across all trees and time series, are reported below.

sycamore

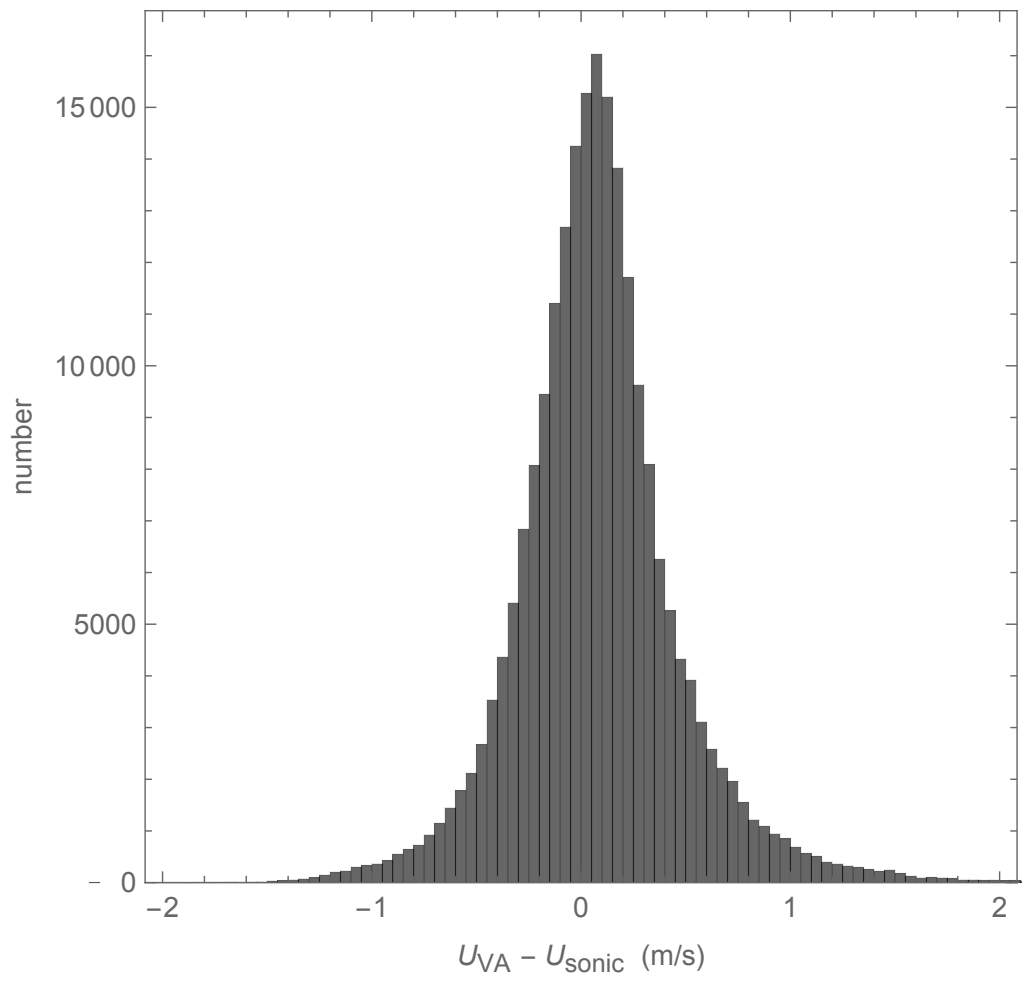

$$\text{mean} = 0.08 \pm 0.40$$

oak

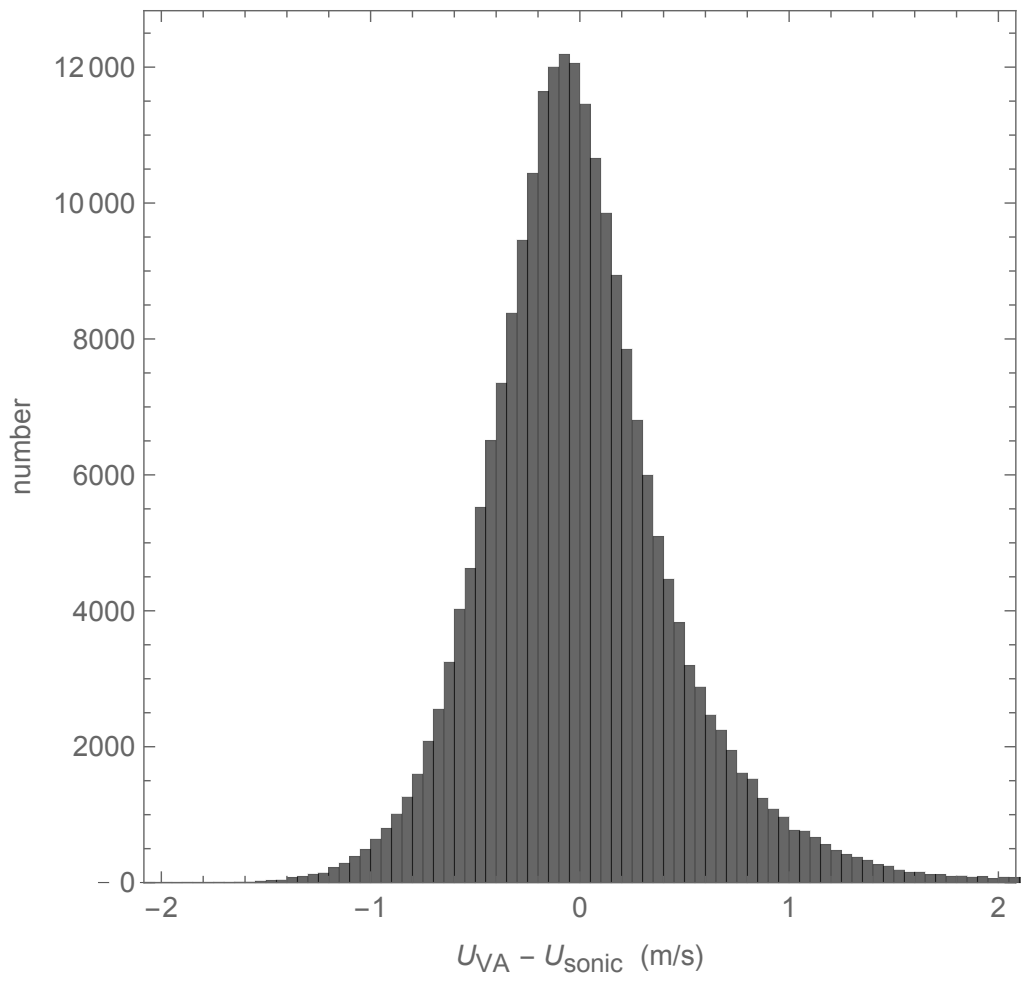

$$\text{mean} = 0.00 \pm 0.47$$

## Section D – Images of the vegetation used in this study

D1. Vegetation measured in the CAST wind tunnel[15]

bullgrass

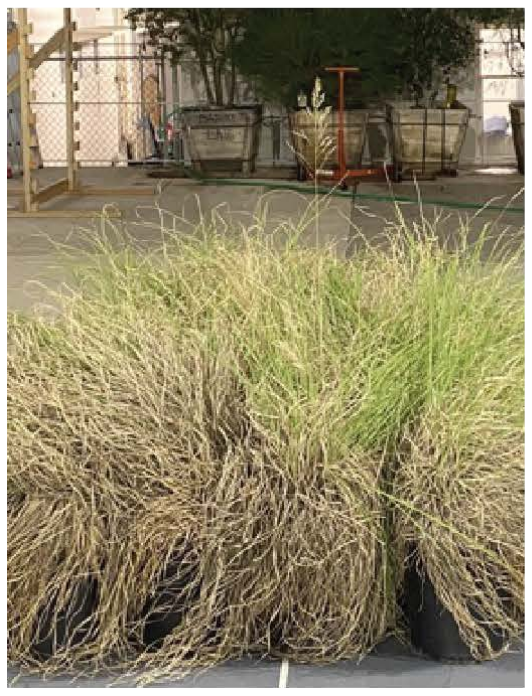

camphor

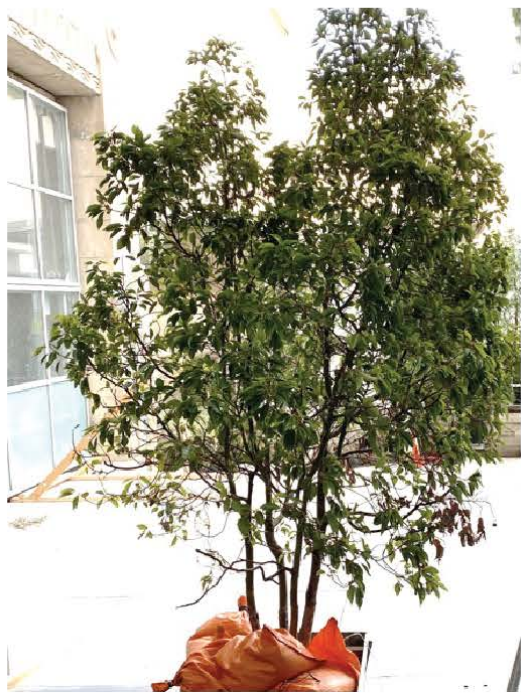

mesquite

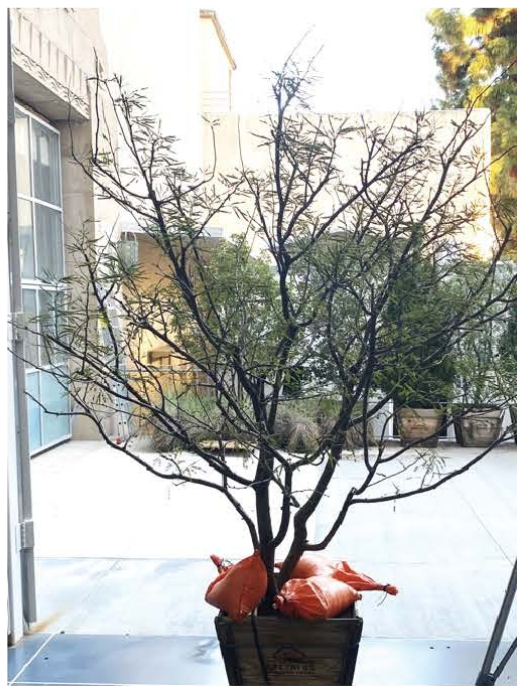

oak

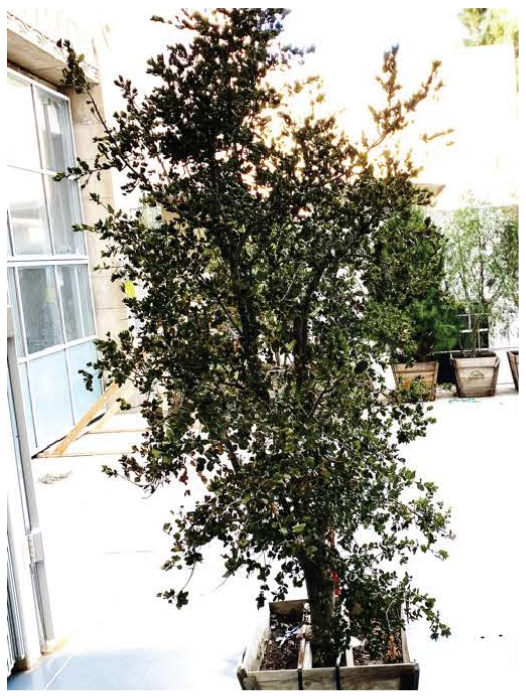

olive

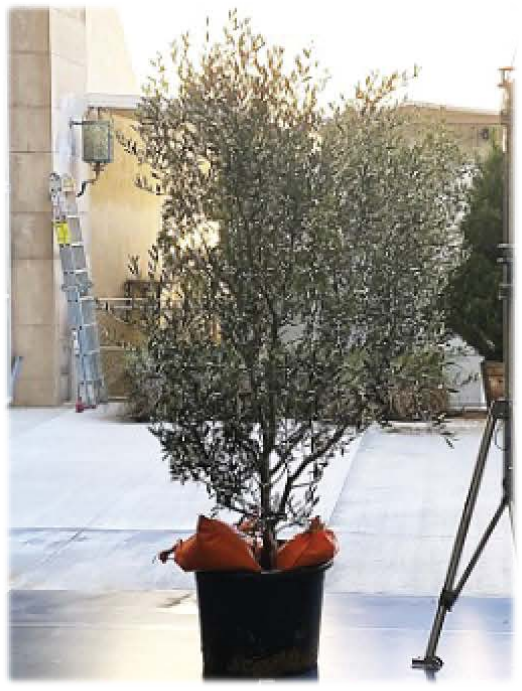

paperbark

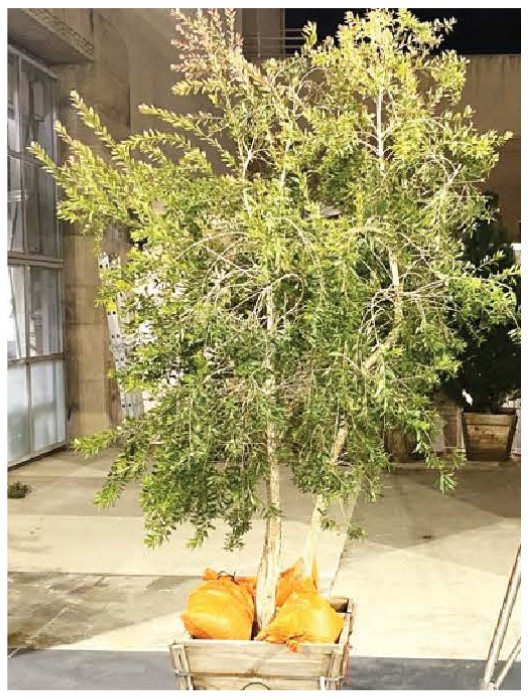

pepper

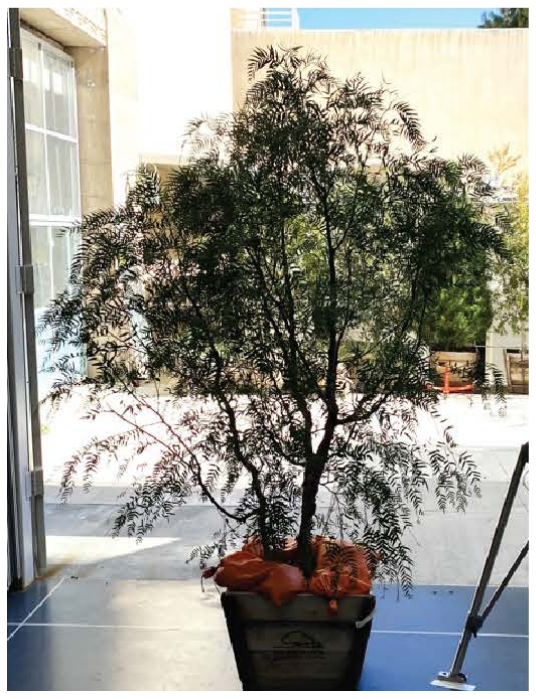

pine

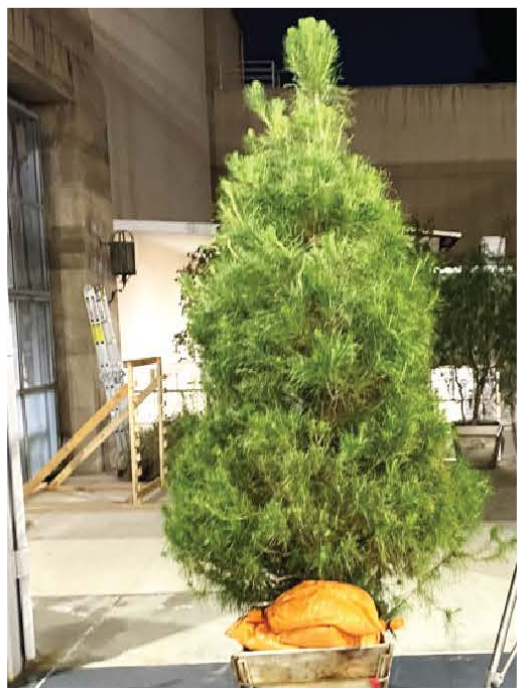

D2. Trees on the Caltech campus

sycamore

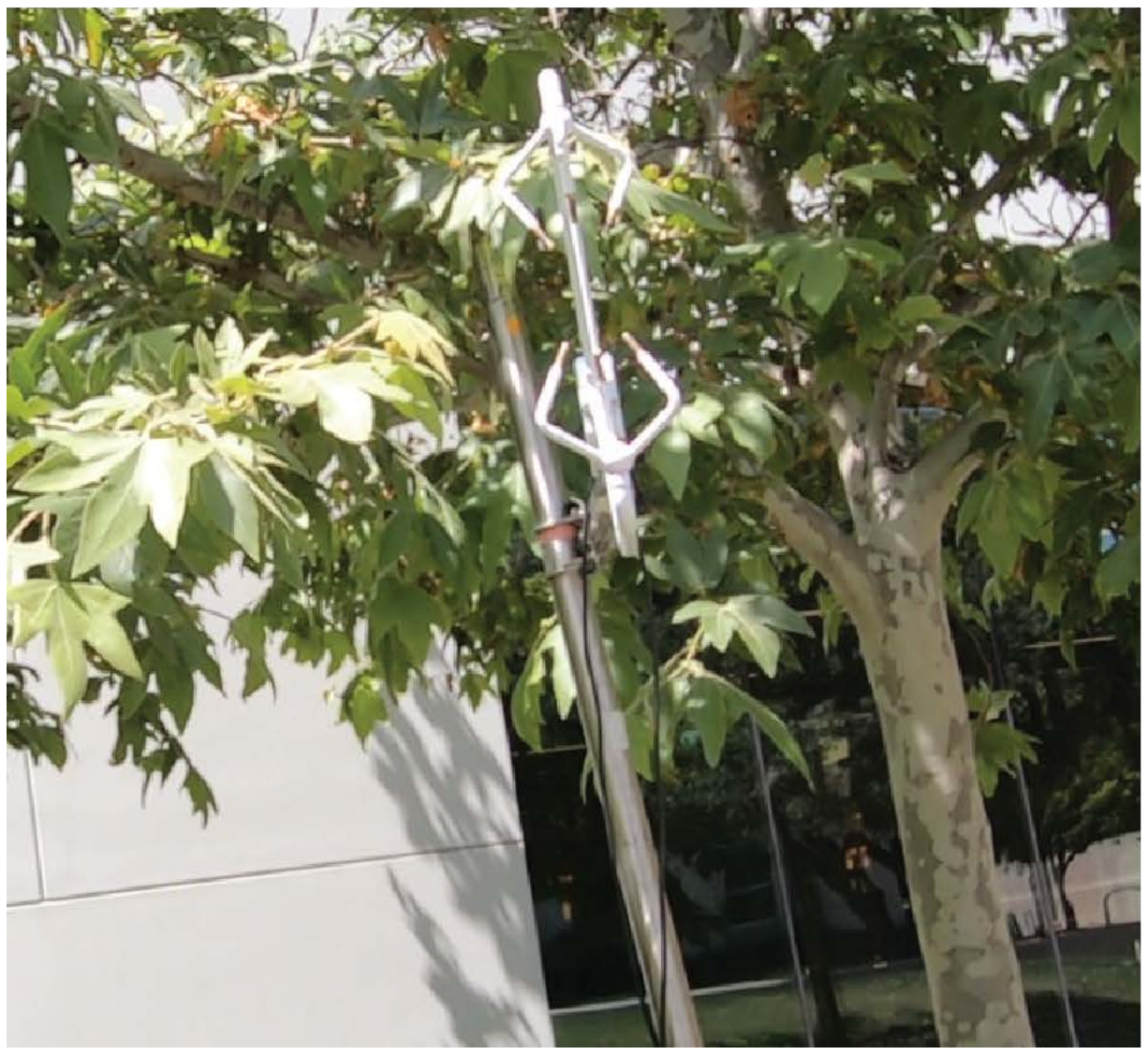

oak

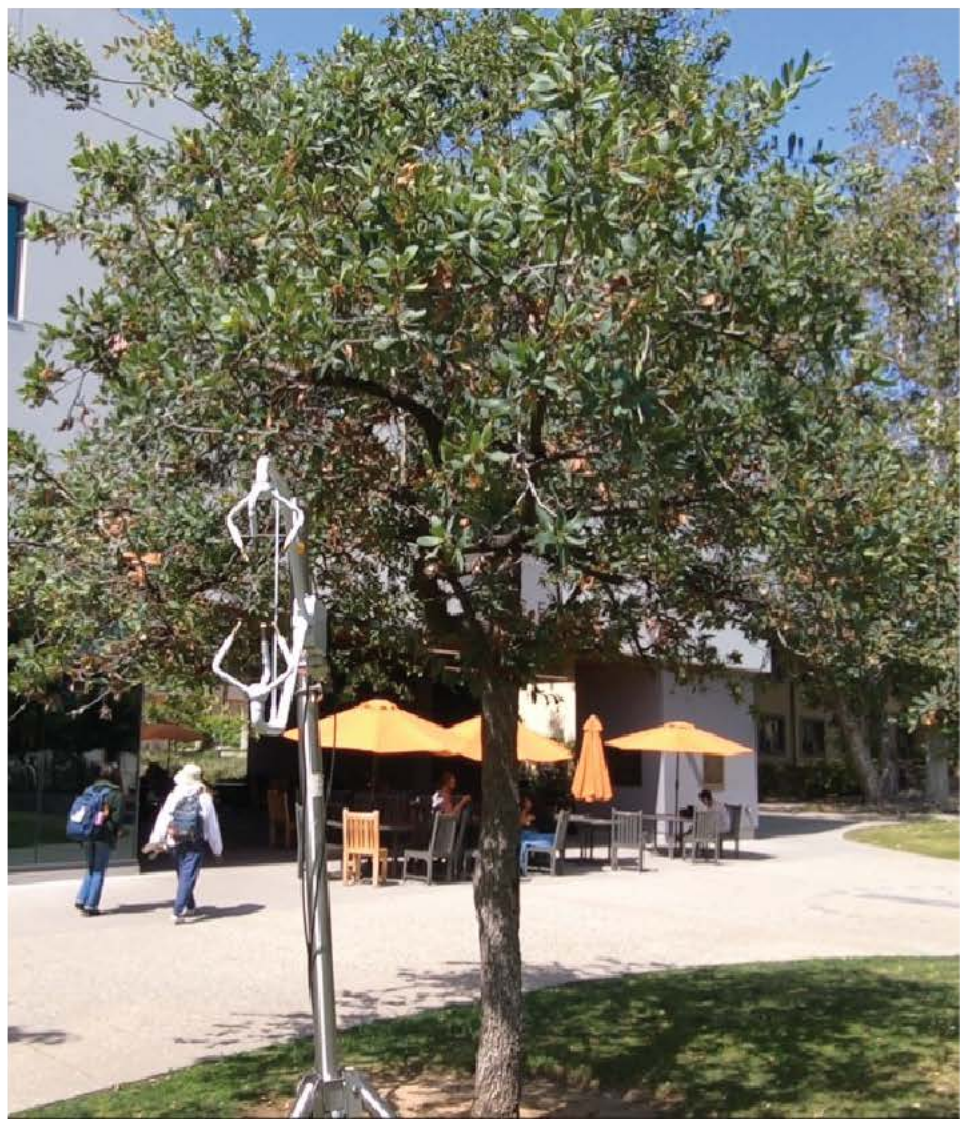